\documentclass[
aps,%
12pt,%
final,%
notitlepage,%
oneside,%
onecolumn,%
nobibnotes,%
nofootinbib,%
superscriptaddress,%
noshowpacs,%
centertags,%
noshowkeys,%
amssymb] {revtex4}
\usepackage{amsmath,graphicx}

\def\,{\ifmmode\mskip\thinmuskip\else\leavevmode\thinspace\fi}

\def\epsilon{\varepsilon}
\def\phi{\varphi}

\newcommand\nn{\nonumber}
\newcommand\dd{\mathrm d}
\newcommand\m[1]{\mathrm {#1}}
\newcommand\Tr{\textmd{Tr\,}}

\begin{document}

\title{The lowest order inelastic QED processes at polarized
photon-electron high energy collisions}

\author{E.~Barto\v{s}}
\altaffiliation{Department of Theoretical Physics,
Comenius University, 84248 Bratislava, Slovakia.}
\affiliation{Joint Institute for Nuclear Research, 141980 Dubna,
Russia}
\author{M.~V.~Galynskii}
\affiliation{Stepanov Institute of Physics BAS, 220072 Minsk,
Skorina ave. 70, Belarus}
\author{S.~R.~Gevorkyan}
\altaffiliation{Yerevan Physics Institute, 375036 Yerevan,
Armenia}
\affiliation{Joint Institute for Nuclear Research, 141980 Dubna,
Russia}
\author{E.~A.~Kuraev}
\affiliation{Joint Institute for Nuclear Research, 141980 Dubna,
Russia}

\date{August 6, 2003}

\begin{abstract}
The compact expressions for cross sections of photoproduction of a
pair of charged particles $\mathrm{e}^+,\mathrm{e}^-$;
$\mu^+,\mu^-$; $\pi^+,\pi^-$ as well as the double Compton
scattering process are given. The explicit analytic expressions
for the case of polarized photon and the initial electron in the
kinematics when all the particles can be considered as a massless
ones are presented. The photon polarization is described in the
terms of Stokes parameters.
\end{abstract}

\maketitle

A process of charged pair photoproduction on electron is usually
used as a polarimeter process. A lot of attention was paid, for
instance, to a trident production process by a linearly polarized
photon (with $\mathrm{e}^-\mathrm{e}^-\mathrm{e}^+$ final state)
where the differential distribution over the recoil electron can
be used to arrange the left-right azimuthal asymmetry of an order
of $14\%$ (see for instance \cite{Aku} and the references
therein).

Linear $\mathrm{e}^+\mathrm{e}^-$ high energy colliders (planned
to be arranged \cite{TESLA}) provide a possibility (using the
backward laser Compton scattering) to obtain the photon-electron
as well as photon-photon colliding beams. The problem of
calibration as well as the problem of important QED background are
to be taken into account for this kind of colliders. The QED
processes at $\gamma \mathrm{e}^\pm$ colliders are relevant for
this purposes. For colliding high energy electron-photon beams
with detections of large-angle emitted final particles the totally
differential cross sections can be useful. For our knowledge the
relevant formulae are absent in the literature,which was the
motivation of our paper. It is organized as follows. First, using
the standard methods \cite{Berends,Galynskii}, we calculate the
chiral amplitudes of processes
\begin{gather}
\gamma(k,\lambda_\gamma)+\m{e}^-(p,\lambda_e)\to
\m{e}^-(p',\lambda_{e})+\gamma(k_1,\lambda_1), \label{eq:1a}\\
\gamma(k,\lambda_\gamma)+\m{e}^-(p,\lambda_e)\to
\m{e}^-(p',\lambda')+a(q_-,\lambda_-)+
\bar{a}(q_+,\lambda_+),\;a=\m{e}^-,\mu^-, \label{eq:1b}\\
\gamma(k,\lambda_\gamma)+\m{e}^-(p,\lambda_e)\to
\m{e}^-(p',\lambda')+\pi^-(q_-)+\pi^+(q_+), \label{eq:1c}\\
\gamma(k,\lambda_\gamma)+\m{e}^-(p,\lambda_e)\to
\m{e}^-(p',\lambda')+\gamma(k_1,\lambda_1)+\gamma(k_2,\lambda_2),
\label{eq:1d}
\end{gather}
with the parameters $\lambda_i=\pm 1$ describe the chiral states
of particles. We consider the experimental setup, when the
polarization of final particles is not measured. Using the
explicit form of chiral amplitudes of the production processes
mentioned above, we construct the chiral matrix. Then, converting
it with photon density matrix, we obtain the cross sections of the
photoproduction processes for arbitrary polarization state of the
initial photon.

In Appendix we give some details of the calculations.

We consider the kinematics when all the 4-vector scalar products
defined by
\begin{gather}
s=2p.p',\quad s_1=2q_-.q_+,\quad t=-2p.q_-,\quad t_1=-2p'.q_+,\\
\nn u=-2p.q_+,\quad  u_1=-2p'.q_-,\quad \chi=2k.p,\quad \chi'=2k.p',\\
\nn \chi_\pm=2k.q_\pm,\quad \chi_j=2k_j.p,\quad
\chi_j'=2k_j.p',\quad j=1,2,
\end{gather}
are large compared with the electron, pion and muon masses squared
\begin{gather}
s\sim s_1\sim -t\sim -t_1\sim -u\sim -u_1\sim \chi_j
\sim\chi_j'\gg m_\mu^2, \\ \nn p^2=p^{{'}2}=q_\pm^2=k^2=k_j^2=0.
\end{gather}
These kinematical invariants are not independent ones. Some
relations can be obtained using the conservation law and mass
shell conditions
\begin{gather}
s_1+t+u=s+t_1+u_1,\quad \chi=s_1-t_1-u_1,\quad \chi'=-s_1-t-u,\\
\nn \chi_+=u_1+s-t,\quad \chi_-=t_1+s-u,\quad
2k_1k_2=\chi-\chi'-s.
\end{gather}

Photon polarization 4-vectors with the definite chirality
$\lambda=\pm1$, $\epsilon_\mu^\lambda(k)$,
$\hat{\epsilon}^\lambda=\gamma^\mu \epsilon_\mu^\lambda$ can be
put in 5 different representations \cite{Berends}
\begin{gather}
\hat{\epsilon}^\lambda_p(k_j)=
N_p(k_i)[\hat{p}'\hat{p}\hat{k}_i\omega_{-\lambda}-
\hat{k}_i\hat{p}'\hat{p}\omega_\lambda],\; k_j=k,k_1,k_2, \nn \\
\hat{\epsilon}^\lambda_q(k)=N_q[\hat{q}_-\hat{q}_+\hat{k}\omega_{-\lambda}-
\hat{k}\hat{q}_-\hat{q}_+\omega_\lambda], \nn \\
\hat{\epsilon}^\lambda_{t1}(k)=N_{t1}[\hat{p}'\hat{q}_+\hat{k}\omega_{-\lambda}-
\hat{k}\hat{p}'\hat{q}_+\omega_\lambda], \\
\hat{\epsilon}^\lambda_t(k)=N_t[\hat{q}_-\hat{p}\hat{k}\omega_{-\lambda}-
\hat{k}\hat{q}_-\hat{p}\omega_\lambda], \nn \\
\epsilon^\lambda_{\pi\mu}(k)=N_q[k.q_-q_{+\mu}-k.q_+q_{-\mu}+
i\lambda\epsilon_{\mu\alpha\beta\gamma}q_-^\alpha q_+^\beta
k^\gamma], \nn
\end{gather}
with
\begin{gather}
\omega_\lambda=(1/2)(1+\lambda\gamma_5),\quad
\omega_\pm^2=\omega_\pm,\quad\omega_+\omega_-=0,\\
\nonumber N=N_p(k)=[s\chi\chi'/2]^{-1/2},\;
N_q=[s_1\chi_+\chi_-/2]^{-1/2},\;N_{t1}=[-t_1\chi'\chi_+/2]^{-1/2},\\
\nn N_{1,2}=N_p(k_{1,2})=[s\chi_{1,2}\chi'_{1,2}/2]^{-1/2},\;
N_t=[-t\chi\chi_-/2]^{-1/2}.
\end{gather}
Here we omit the terms of type $\hat{k},k_\mu$ in the right-hand
side of equations for $\epsilon(k)$, which are irrelevant due to
gauge conditions. These vectors obey the conditions
\begin{gather}
\epsilon_p^\lambda(k_j)=\epsilon_j^\lambda
\m{e}^{j\lambda\phi_j}+c_jk,\quad \m{e}^{i\lambda\phi_j}=
-(1/4)\Tr \hat{\epsilon}_p^{\lambda}\hat{\epsilon}_j^{-\lambda},\\
\nn \epsilon_i^\lambda \epsilon_j^\lambda=0\;\;(i\ne j,\;
i=j),\quad \epsilon_i^\lambda \epsilon_j^{-\lambda}=-1\;\; (i=j).
\end{gather}
The quantity $c_i$ can be safely omitted due to gauge invariance
of the amplitude. We will systematically omit such terms.
Performing the calculations of chiral amplitudes we use
$\epsilon^\lambda_p$ and transform it to the relevant form
considering the definite Feynman amplitudes. Note as well the
useful relation \cite{Berends}
\begin{eqnarray}\label{eq:use}
sN_p\hat{\epsilon}^\lambda_p+s_1N_q\hat{\epsilon}^\lambda_q=
-[tN_{t1}\hat{\epsilon}^\lambda_1+
t_1N_t\hat{\epsilon}^\lambda_2].
\end{eqnarray}
The relation (\ref{eq:use}) can be written in the form
\begin{gather} \label{eq:es}
[s_\pm]=-\m{e}^{\pm i\phi_2}[t_\pm],\quad
[s_\pm]=sN_p+s_1N_q\m{e}^{\pm i\phi_q},\\ \nn
[t_{\pm}]=N_tt_1+tN_{t1}\m{e}^{\pm i\phi_{12}},\quad
\epsilon_2^\lambda=\epsilon_1^\lambda \m{e}^{i\lambda\phi_{12}}.
\end{gather}
Following \cite{Berends} one can obtain (see Appendix)
\begin{gather}
[t_+][t_-]=[s_+][s_-]=\frac{W}{2}=r_s+r_{s_1}+r_t+r_{t_1}-r_u-r_{u_1},
\\ \nn W=-\bigg(\frac{q_+}{k.q_+}-\frac{q_-}{k.q_-}+
\frac{p}{k.p}-\frac{p'}{k.p'}\bigg)^2,
\\ \nn r_s=\frac{2s}{\chi\chi'},\,
r_{s_1}=\frac{2s_1}{\chi_+\chi_-},\, r_t=\frac{-2t}{\chi\chi_-},\,
r_{t_1}=\frac{-2t_1}{\chi'\chi_+},\, r_u=\frac{-2u}{\chi\chi_+},\,
r_{u_1}=\frac{-2u_1}{\chi'\chi_-}\:\cdot
\end{gather}
Chiral states of leptons are defined by
\begin{gather}
u^\lambda(p)=\omega_\lambda u(p),\quad
\bar{u}(p)^\lambda=\bar{u}(p)\omega_{-\lambda}, \\ \nonumber
v^\lambda(p)=\omega_{-\lambda}v(p),\quad
\bar{v}^\lambda(p)=\bar{v}(p)\omega_\lambda, \\ \nonumber
u^\lambda(p)\bar{u}^\lambda(p)=\omega_\lambda\hat{p},\quad
v^\lambda(p)\bar{v}^\lambda(p)=\omega_{-\lambda}\hat{p}.
\end{gather}
The spin density matrix for the initial electron in the
ultrarelativistic case has the form
\begin{eqnarray}
u(p,a)\bar{u}(p,a)=\frac{1}{2}\hat{p}[1-\gamma_5(\lambda+
\hat{\lambda}_\bot)],\quad\lambda^2+|\lambda_\bot|^2\le 1,\quad
-1<\lambda<1.
\end{eqnarray}
The effect of transversal polarization of the fermion
$\lambda_\bot$ is negligible (the relevant contribution is
suppressed by a factor $m/\sqrt{\chi}$ compared to the case of
longitudinal polarization of fermions). Cross section for the
incomplete polarization of the initial electron will have the form
\begin{eqnarray}
\dd\sigma=\zeta_+\dd\sigma_++\zeta_-\dd\sigma_-,\quad
\zeta_\pm=\frac{1}{2}(1\pm\lambda)
\end{eqnarray}
and $\dd\sigma_\pm$ corresponds to the case of complete
longitudinal polarization of the initial electron described by the
pure chiral state. Below we suggest that the initial electron
belongs to the pure chiral state.

The chiral amplitudes
$M^{\lambda_-\lambda_+\lambda'}_{\lambda_\gamma\lambda}$ for
processes of a charged pair of lepton production,
$M^{\lambda'}_{\lambda_\gamma\lambda}$ for a pair of charged pion
production and
$M^{\lambda_1\lambda_2\lambda'}_{\lambda_\gamma\lambda}$ for a two
photon-electron final states are defined as a usual matrix element
calculated with chiral states of photons and leptons.

The matrix elements in the lower order of perturbation theory are
determined by tree-type Feynman diagrams.

The matrix element for muon pair production has the form
$M_{\mu^+\mu^-}=M_1+M_2$, with
\begin{gather}
M_1=-\frac{1}{s}\bar{u}(q_-)\bigg[\hat{\epsilon}\frac{\hat{q}_-
-\hat{k}}{-\chi_-}\gamma_\eta+ \gamma_\eta
\frac{-\hat{q}_++\hat{k}}{-\chi_+}\hat{\epsilon}\bigg]v(q_+)
\bar{u}(p')\gamma_\eta u(p), \\ \nonumber
M_2=\frac{1}{s_1}\bar{u}(q_-)\gamma_\sigma v(q_+) \bar{u}(p')
\bigg[\hat{\epsilon}\frac{\hat{p}'-\hat{k}}{-\chi'}\gamma_\sigma+
\gamma_\sigma\frac{\hat{p}+\hat{k}}{\chi}\hat{\epsilon}\bigg]u(p).
\end{gather}
For the $\m{e}^+\m{e}^-$ pair production the identity of electrons
must be taken into account which result in the appearance of two
additional terms
\begin{eqnarray}
M_{\m{e}^+\m{e}^-}=M_1+M_2-\tilde{M}_1-\tilde{M}_2,
\end{eqnarray}
with
\begin{gather}
\tilde{M}_1=\frac{1}{t}\bar{u}(q_-)\gamma_\rho u(p)
\bar{u}(p')\bigg[\hat{\epsilon}\frac{\hat{p}'-\hat{k}}
{-\chi'}\gamma_\rho+\gamma_\rho
\frac{-\hat{q}_++\hat{k}}{-\chi_+}\hat{\epsilon}\bigg]v(q_+), \\
\nonumber
\tilde{M}_2=-\frac{1}{t_1}\bar{u}(q_-)\bigg[\hat{\epsilon}
\frac{\hat{q}_--\hat{k}}{-\chi_-}\gamma_\eta+\gamma_\eta
\frac{\hat{p}+\hat{k}}{\chi}\hat{\epsilon}\bigg]u(p)
\bar{u}(p')\gamma_\eta v(q_+).
\end{gather}
For the $\pi^+\pi^-$ pair we have $M_{\pi^+\pi^-}=M_3+M_4$ with
\begin{align}
M_3=&-\frac{1}{s}\bar{u}(p')\gamma_\eta u(p)\\ \nn
&\times\bigg[-2(q_--q_+-k)_\eta \frac{\hat{q}_-\hat{e}} {\chi_-}+
2(q_--q_++k)_\eta\frac{\hat{q}_+\hat{\epsilon}}{\chi_+}-2e_\eta\bigg],
\\ \nonumber
M_4=&\frac{1}{s_1}(q_--q_+)_\sigma\bar{u}(p')
\bigg[\gamma_\sigma\frac{\hat{p}+\hat{k}}{\chi}\hat{\epsilon}+
\hat{\epsilon}\frac{\hat{p}'-\hat{k}}{-\chi'}\gamma_\sigma\bigg]u(p).
\end{align}
Using these expressions we calculate the chiral amplitudes of the
processes.

The matrix elements squared summed over the spin states of the
final particles have the form of the conversion of the chiral
matrix with the photon density matrix
\begin{equation}\sum |M|^2=\frac{1}{2\,} \Tr
\begin{pmatrix}
 m_{11} & m_{12} \\  m_{21} & m_{22}
\end{pmatrix}
\begin{pmatrix}
 1+\xi_2 & i\xi_1-\xi_3 \\ -i\xi_1-\xi_3  & 1-\xi_2
\end{pmatrix},
\end{equation}
with the vector of photon polarization
$\vec{\xi}=(\xi_1,\xi_2,\xi_3)$ parameterized by Stokes parameters
fulfilling the condition  $\xi_1^2+\xi_2^2+\xi_3^2\le 1$.

The matrix elements of the chiral matrix $m_{ij}$ are constructed
from the chiral amplitudes of the process
$M^{\lambda_-\lambda_+\lambda'}_{\lambda_\gamma\lambda_e}$ as
\begin{gather}
m_{11}=\sum_{\lambda_-\lambda_+\lambda '}
\big|M^{\lambda_-\lambda_+\lambda '}_{++}\big|^2,\quad
m_{22}=\sum_{\lambda_-\lambda_+\lambda '}
\big|M^{\lambda_-\lambda_+\lambda '}_{-+}\big|^2, \nn\\
m_{12}=\sum_{\lambda_-\lambda_+\lambda '}
M^{\lambda_-\lambda_+\lambda'}_{++}
\big(M^{\lambda_-\lambda_+\lambda'}_{-+}\big)^*,\quad
m_{21}=m_{12}^*.
\end{gather}

We put here only half of all chiral amplitudes which correspond to
$\lambda_e=+1$. The other half can be obtained from these ones by
a space parity operation (replacement $\omega_\lambda$ by
$\omega_{-\lambda}$ and $[s_+]$ by $[s_-]$).

For the process of Compton scattering the matrix element
$M_{\lambda_e}^{\lambda_\gamma\lambda_1}$ has the form (here and
further we omit the factor $i(4\pi\alpha)^{n/2}$)
\begin{eqnarray}
M^{+-}_{+}=sN_1N\chi\bar{u}(p')\hat{k}\omega_+u(p),\quad
M^{-+}_{+}=sN_1N\chi'\bar{u}(p')\hat{k}\omega_+u(p).
\end{eqnarray}

For the muon pair production for
$M_{\lambda_\gamma,\lambda}^{\lambda_-\lambda_+\lambda'}$ we have
\begin{align}
M^{+-+}_{++}&=-\frac{2u}{s_1}[s_+]\frac{\bar{u}(q_-)\hat{q}_+
\omega_+u(p)}{\bar{v}(q_+)\hat{p}\omega_+u(p')},\\ \nn
M^{+-+}_{-+}&=\frac{2u_1}{s}[s_-]\frac{\bar{u}(q_-)\hat{p}'
\omega_+u(p)}{\bar{v}(q_+)\hat{q}_-\omega_+u(p')},\\ \nonumber
M^{+--}_{+-}&=-\frac{2t_1}{ss_1}[s_+]\frac{\bar{u}(q_-)\hat{q}_+
\hat{p}'\omega_-u(p)}{\bar{v}(q_+)\omega_-u(p')},\\\nn
M^{+--}_{--}&=-\frac{2}{ss_1}[s_-]\frac{\bar{u}(q_-)\hat{p}\hat{p}'
\hat{q}_+\hat{q}_-\omega_-u(p)}{\bar{v}(q_+)\omega_-u(p')}.
\end{align}

For the case of the $\m{e}^+\m{e}^-$ pair production
\begin{align}
M^{+--}_{+-}&=-\frac{2t_1}{ss_1}[s_+]\frac{\bar{u}(q_-)\hat{q}_+
\hat{p}_1'\omega_-u(p)}{\bar{v}(q_+)\omega_-u(p')},\\ \nn
M^{+--}_{+-}&=-\frac{2}{ss_1}[s_-]\frac{\bar{u}(q_-)\hat{p}_1\hat{p}_1'
\hat{q}_+\hat{q}_-\omega_-u(p)}{\bar{v}(q_+)\omega_-u(p')},\\\nn
M^{++-}_{++}&=-\frac{2s}{tt_1}[s_+]\frac{\bar{u}(q_-)\hat{p}_1
\hat{p}_1'\omega_-v(q_+)}{\bar{u}(p)\omega_-u(p')},\\\nn
M^{++-}_{-+}&=-\frac{2}{tt_1}[s_-]\frac{\bar{u}(q_-)\hat{q}_+\hat{p}_1'
\hat{p}_1\hat{q}_-\omega_-v(q_+)}{\bar{u}(p)\omega_-u(p')},\\\nn
M^{+-+}_{++}&=\frac{2[s_+]}{tt_1ss_1}[A_+-ss_1]\bar{u}(p')\hat{q}_+
\omega_+u(p)\bar{u}(q_-)\hat{p}_1\omega_+v(q_+),\\\nn
M^{+-+}_{-+}&=\frac{2[s_-]}{tt_1ss_1}[A_--ss_1]\bar{u}(p')\hat{q}_-
\omega_+u(p)\bar{u}(q_-)\hat{p}_1'\omega_+v(q_+),
\end{align}
with $A_\pm=\Tr\hat{q}_+\hat{p}_1'\hat{p}_1\hat{q}_-\omega_\pm$.

For process of charged pion pair production
$M^{\lambda'}_{\lambda_\gamma\lambda}$ we have
\begin{align}
&M^+_{++}=\frac{2}{ss_1}[s_+]\bar{u}(p')
(u\hat{q}_--t\hat{q}_+)\omega_+u(p), \\ \nn
&M^+_{-+}=\frac{2}{ss_1}[s_-]\bar{u}(p')
(u_1\hat{q}_+-t_1\hat{q}_-)\omega_+u(p).
\end{align}

For the case of double Compton scattering chiral amplitudes
$M_{\lambda_\gamma,\lambda}^{\lambda_1\lambda_2\lambda'}$ we have
\begin{align}
&M^{+++}_{-+}=-s^2\chi N_1N_2N\bar{u}(p')\hat{k}\omega_+u(p),\\
\nonumber &M^{-++}_{++}=-s^2\chi_1
N_1N_2N\bar{u}(p')\hat{k}_1\omega_+u(p),\\ \nonumber
&M^{+-+}_{++}=-s^2\chi_2
N_1N_2N\bar{u}(p')\hat{k}_2\omega_+u(p),\\ \nonumber
&M^{--+}_{++}=-s^2\chi' N_1N_2N\bar{u}(p')\hat{k}\omega_+u(p),\\
\nonumber &M^{+-+}_{-+}=-s^2\chi_1'
N_1N_2N\bar{u}(p')\hat{k}_1\omega_+u(p),\\ \nonumber
&M^{-++}_{-+}=-s^2\chi_2' N_1N_2N\bar{u}(p')\hat{k}_2\omega_+u(p).
\end{align}
The remaining amplitudes are equal to zero. We put here some
useful relations
\begin{gather}
\bigg|\frac{\bar{u}(q_-)\hat{q}_+\omega_+u(p)}
{\bar{v}(q_+)\hat{p}_1\omega_+u(p')}\bigg|^2=\frac{s_1}{s},\quad
\quad \bigg|\frac{\bar{u}(q_-)\hat{p}_1'\omega_+u(p)}
{\bar{v}(q_+)\hat{q}_-\omega_+u(p')}\bigg|^2=\frac{s}{s_1}, \quad
\quad\\\nn
\bigg|\frac{\bar{u}(q_-)\hat{q}_+\hat{p}_1'\omega_-u(p)}
{\bar{v}(q_+)\omega_-u(p')}\bigg|^2=ss_1,\quad
\bigg|\frac{\bar{u}(q_-)\hat{p}_1\hat{p}_1'\hat{q}_+\hat{q}_-
\omega_-u(p)}{\bar{v}(q_+)\omega_-u(p')}\bigg|^2=ss_1t^2,\\\nn
\bigg|\frac{\bar{u}(q_-)\hat{p}_1\hat{p}_1'\omega_-v(q_+)}
{\bar{u}(p)\omega_-u(p')}\bigg|^2=tt_1,\quad
\bigg|\frac{\bar{u}(q_-)\hat{q}_+\hat{p}_1'\hat{p}_1\hat{q}_-
\omega_-v(q_+)}{\bar{u}(p)\omega_-u(p')}\bigg|^2=s_1^2tt_1,
\end{gather}
\begin{gather}
|\bar{u}(p')\hat{q}_+\omega_+u(p)\bar{u}(q_-)\hat{p}_1
\omega_+v(q_+)|^2=u^2tt_1, \nn\\ \nn
|\bar{u}(p')\hat{q}_-\omega_+u(p)\bar{u}(q_-)
\hat{p}_1'\omega_+v(q_+)|^2=u_1^2tt_1,\\\nn
|\bar{u}(p')(u\hat{q}_--t\hat{q}_+)\omega_+u(p)|^2=ss_1tu,\\\nn
|\bar{u}(p')(u_1\hat{q}_+-t_1\hat{q}_-)\omega_+u(p)|^2=
ss_1t_1u_1, \\ \nonumber
|\bar{u}(p')\hat{k}\omega_+u(p)|^2=\chi\chi',\quad
|\bar{u}(p')\hat{k}_1\omega_+u(p)|^2=\chi_1\chi_1',
\\\nonumber
|\bar{u}(p')\hat{k}_2\omega_+u(p)|^2=\chi_2\chi_2'.
\end{gather}
Besides we have
$|A_\pm|^2=|\bar{u}(q_+)\hat{p}_1'\hat{p}_1\hat{q}_-\omega_\pm
u(q_+)|^2=ss_1tt_1$ and $|A_\pm-ss_1|^2=ss_1uu_1$.

For the case of muon pair production we have
\begin{align}
m_{11}=&4[s_+][s_-]\frac{u^2+t^2}{ss_1},\quad m_{22}=4[s_+][s_-]
\frac{u_1^2+t_1^2}{ss_1},\\ \nn
m_{12}=&-\frac{4[s_+]^2}{(ss_1)^2}\Big[(ss_1)^2+(tt_1)^2+(uu_1)^2-
2tt_1uu_1-ss_1(tt_1+uu_1)\\\nn &+4i(tt_1-uu_1)A\Big],
\end{align}
where $A=\epsilon_{\mu\nu\rho\sigma}q_+^\mu q_-^\nu p^\rho
{p'}^\sigma$.

For the electron pair production we have
\begin{align}
m_{11}=&\frac{4[s_+][s_-]}{ss_1tt_1}\Big[t^3t_1+u^3u_1+s^3s_1\Big],\quad
m_{22}=\frac{4[s_+][s_-]}{ss_1tt_1}\Big[t_1^3t+u_1^3u+s_1^3s\Big],\\\nn
m_{12}=&\frac{4[s_+]^2}{(ss_1tt_1)^2}\Big[(uu_1)^2-(ss_1)^2-(tt_1)^2\Big]
\Big[\frac{1}{2}((uu_1)^2+(ss_1)^2+(tt_1)^2\\\nn
&-2uu_1(ss_1+tt_1))+ 2iA(uu_1-ss_1-tt_1)\Big].
\end{align}

For pions one gets
\begin{align}
m_{11}=&[s_+][s_-]tu,\quad m_{22}=[s_+][s_-]t_1u_1,\\\nn
m_{12}=&\frac{[s_+]^2}{ss_1}\Big[\frac{1}{2}(uu_1-tt_1)^2
-ss_1(uu_1+tt_1)+2i(tt_1-uu_1)A\Big].
\end{align}
The explicit expression for $[s_+]^2$ is given in Appendix.

For the double Compton scattering we have
\begin{align}
m_{11}=&\frac{8s}{D}\Big[\chi_1^3\chi_1'+\chi_2^3\chi_2'+
{\chi'}^3\chi\Big],\quad
m_{22}=\frac{8s}{D}\Big[{\chi_1'}^3\chi_1+{\chi_2'}^3\chi_2+
{\chi}^3\chi'\Big],\nn\\
m_{12}=&\frac{8s}{D}\bigg\{\frac{1}{2}(\chi_1\chi_2'+\chi_2\chi_1')
\Big[\chi_1\chi_2'+\chi_2\chi_1'+s(s+\chi'-\chi)\Big]\\ \nn
&+2iB(\chi_2\chi_1'-\chi_1\chi_2')\bigg\},
\end{align}
with $D=\chi\chi'\chi_1\chi_1'\chi_2\chi_2'$ and
$B=\epsilon_{\mu\nu\rho\sigma}k_2^\mu k_1^\nu p^\rho {p'}^\sigma$.

The differential cross section of the Compton scattering process
(\ref{eq:1a}) has the form
\begin{eqnarray}
\dd\sigma_{\lambda_\m{e}\vec{\xi}}^{\gamma \m{e}^-\to\gamma
\m{e}^-}= \frac{\alpha^2}{4\chi}
\bigg[\frac{\chi^2+{\chi'}^2}{\chi\chi'}+ \xi_2\lambda_e
\frac{\chi^2-{\chi'}^2}{\chi\chi'}\bigg]\dd O_\gamma .
\end{eqnarray}

The cross section of processes (\ref{eq:1b}, \ref{eq:1c}) has the
form
\begin{gather}\label{}
\frac{\dd\sigma}{\dd\Gamma}=\frac{\alpha^3}{2\pi^2\chi}
\Big[m_{11}+m_{22}+\xi_2\lambda_e(m_{11}-m_{22})
-2\xi_3\mathrm{Re}
(m_{12})+2\xi_1\mathrm{Im} (m_{12})\Big],\nn\\
\dd\Gamma=\frac{\dd^3p'}{\epsilon'}\frac{\dd^3q_-}{\epsilon_-}
\frac{\dd^3q_+}{\epsilon_+}\delta^4(p+k-p'-q_+-q_-)
\end{gather}

The cross section of processes (\ref{eq:1d}) has the form
\begin{align}\label{}
\label{11}
\frac{\dd\sigma}{\dd\Gamma_\gamma}=&\frac{1}{2!}\frac{\alpha^3s}
{2\pi^2\chi D} \Big[\chi\chi' ({\chi}^2+{\chi'}^{2})+
\chi_1\chi_1'(\chi_1^2+{\chi_1'}^2)+\chi_2\chi_2'(\chi_2^2+{\chi_2'}^2)
 \nn \\
&+4\xi_1B(\chi_1\chi_2'-\chi_2\chi_1')-\xi_3(\chi_1\chi_2'+
\chi_2\chi_1')(\chi\chi'-\chi_1\chi_1'-\chi_2\chi_2') \nn \\
&+\lambda_e\xi_2[\chi\chi'{(\chi'}^2-\chi^2)+\chi_1\chi_1'(\chi_1^2-
{\chi_1'}^2)+\chi_2\chi_2'(\chi_2^2-{\chi_2'}^2)\Big],\\\nn
&\dd\Gamma_\gamma=\frac{\dd^3p'}{\epsilon'}\frac{\dd^3k_1}{\epsilon_1}
\frac{\dd^3k_2}{\epsilon_2}\delta^4(p+k-p'-k_1-k_2),
\end{align}
where the multiplier $1/2!$ takes into account the identity of
photons in the final state.

As a conclusion we note that the unpolarized part of the cross
section dominates over the polarized one, providing the positivity
of cross sections. Asymmetries (the ratios of the polarized part
to the unpolarized one) in general kinematics are the quantities
of an order of unity.

\section*{Appendix}

It is convenient to write down $[s_+]$ (\ref{eq:es}) in the form
\begin{equation}\label{}
[s_+]=\frac{1}{2\sqrt{r_s}}\bigg[2r_s-r_u-r_{u_1}+r_t+r_{t_1}-
8i\frac{(\chi_++\chi_-)A}{\chi\chi'\chi_+\chi_-}\bigg].
\end{equation}
Then the quantity $[s_+]^2$ which enters into the cross sections
of processes (\ref{eq:1b}, \ref{eq:1c}) has the form
\begin{gather}
[s_+]^2=T_1-iT_2A, \\ \nn T_1=r_s-r_{s_1}+r_t+r_{t_1}-r_u-r_{u_1}+
\frac{1}{2r_s}(r_u+r_{u_1}-r_t-r_{t_1})^2, \\ \nn
T_2=\frac{4(\chi_++\chi_-)}{\chi_+\chi_-\chi\chi'r_s}
[2r_s-r_u-r_{u_1}+r_t+r_{t_1}].
\end{gather}
To check the expression $[s_+]^2[s_-]^2=\frac{W^2}{4}$ we used the
relation
$$(\chi_++\chi_-)A=\epsilon_{\mu\nu\rho\sigma}(\chi_+k^\mu q_-^\nu
p^\rho {p'}^\sigma+\chi_-q_+^\mu k^\nu p^\rho {p'}^\sigma).$$

\begin{acknowledgments}
We are grateful to ECT Trento Centrum and to the Institute of
Physics SAS (Bratislava), where part of this work was done. The
work was supported by RFFI grant No. 03--02--17077, INTAS grant
No. 00366 and Slovak Grant Agency for Sciences, grant  No.
2/1111/23. One of us (M.G.) is also grateful to grant BRFFI No.
F03-183.
\end{acknowledgments}

\end{document}